\documentclass[preprint,review,12pt]{elsarticle}

\usepackage{amssymb}

\journal{Physica A: Statistical Mechanics and its Applications}

\begin{document}

\begin{frontmatter}



\title{Explicit Solutions to the Continuous Time Albert-Barab\'asi Scale-Free Model.\tnoteref{t1}}

\tnotetext[t1]{The author would like to thank Sidney Redner for his suggestions that broadened the discussion and intellectual merit of the paper.}


\author[brm]{Benjamin R. Morin\fnref{f1}}
\ead{brmorin@asu.edu}
\address[brm]{School of Human Evolution and Social Change, Arizona State University, Tempe, AZ 85287}
\fntext[f1]{The author is also a member of the Mathematical and Computational Modeling Sciences Center, Arizona State University, PO Box 871904, Tempe, AZ, 85287.}
\begin{abstract}
Presented here is the explicit solution to the continuous time approximation of the Albert-Barab\'asi scale-free network model at any time $t$.  The solution is found by recursively solving the differential equations via integrating factors, identifying a pattern for the coefficients and then proving this observed pattern using induction.
\end{abstract}

\begin{keyword}
Scale-free network \sep Differential equations \sep Inductive solution

\end{keyword}

\end{frontmatter}


\section{Introduction}
\label{Intro}
The Albert-Barab\'asi scale-free network model has received a great deal of attention for well over a decade since the preliminary work, focused on determining the mechanism that producing power-law degree distribution of certain real-world networks, was written \cite{barabasi1999emergence}.  Work rapidly ensued in the application of such network structure to phenomena such as personal preferences, language structure, statistical physics (Bose-Einstein condensation), epidemics and a veritable wealth of biological examples \cite{bianconi2001bose, caldarelli2007scale, chowell28worst, christakis2007spread, lewis2008tastes, PhysRevE.67.031911, PhysRevE.65.065102, PhysRevE.66.065103, pastor2001epidemic, small2007scale}.  The intrinsic concept of preferential attachment, previosly called cumulative advantage, was first proposed by Price as an application to citation networks with a slightly more general model \cite{price1965networks}.  Despite the decades that this concept has persisted in the literature, an explicit solution to the {\em degree distribution} of such a network has been absent.\\

One such method used to describe the dynamic evolution of such a network is the  doubly indexed, by degree and time, difference equation \cite{newman2006structure}
$$p_{k,t}(t_i)=\frac{k-1}{2(t-1)}p_{k-1,t-1}(t_i)+\left(1-\frac{k}{2(t-1)}\right)p_{k,t-1}(t_i).$$  
In this model new nodes are introduced at unit times with a single edge to connect preferentially to highly connected existing nodes.  The solution to this model was found asymptotically, i.e. as $t\rightarrow\infty$, by Dorogovtsev {\em et al.} \cite{dorogovtsev2000structure}.  The continuous time analog was simultaneously solved by Krapivsky {\em et al.}, again in the ``most interesting asymptotic regime ($t\rightarrow\infty$)" \cite{krapivsky2000connectivity}.  It is unclear if Karpivsky and his co-authors found the explicit time $t$ solution for this paper specifically.  Similar later work by Krapivsky and Redner does indeed calculate the explicit solution for the degree distribution for a model where rewiring of connections and a more general preferential structure is modeled \cite{krapivsky2002finiteness}.  In this, and a previous paper (\cite{krapivsky2001organization}), the author's investigate various correlations, e.g. between age of nodes and their connectivity, in addition to utilizing rate equations (differential equations) and the generating function (partial differential equations) to compute the form and various moments of the degree distribution.\\

Exact solutions for the robust metrics (e.g. degree distribution, mean path length, eigenvalue structure, etc...) are important in the analysis of the structure of the networks themselves or phenomena spreading on them.  Indeed, Dorogovtsev {\em et al.} explicitly calculated the path length of a small world network structure, \cite{dorogovtsev2000exactly}, from the discrete model, and Newman {\em et al.} did the same to the mean field model \cite{newman2000mean}.  Wang {\em et al.} address epidemic spread on complex networks without addressing the specific network archetype considered \cite{wang2003epidemic}.  Instead, they analyze the epidemic threshold produced via the eigenvalue structure of complex topology.  Valente has proposed a technique for investigating thresholds of spreads on social networks (either established or as they form) based on the type of adoption employed in the diffusion of innovation by individuals \cite{valente1996social}.  While he does not explicitly address the network topology itself,  it is clear that a need for the distribution on the number of connections, as well as their ``quality", at any time during the spread is important to understanding the resultant diffusion.\\

Presented here is something of a side-generalization of the scale-free model that was previously solved asymptotically by Krapivsky {\em et al.}, again explicitly by Krapivsky and Redner and in other approximate forms by Barab\'asi, Albert and Jeong (\cite{barabasi1999mean}):
\begin{equation}
\dot{N}_k(t)=\Lambda\delta_{k,m}+\Lambda m\left[\pi_{k-1}(t)N_{k-1}(t)(1-\delta_{k,m})-\pi_k(t)N_k(t)\right].
\end{equation}
I allow new nodes to enter the network at constant rate $\Lambda$,  and connect to $m$ existing nodes with density dependent preference $\pi_k(t)=\frac{k}{D(t)}$, where $D(t)=2\Lambda mt+D_0$ is twice the count of edges within the network with $D_0/2$ initial edges.  This is the so-called rate equation(s) for the counts of nodes with degree $k$.  The derivation of such an equation is well handled in a number of review papers \cite{albert2002statistical, boccaletti2006complex, newman2003structure}.  When considering the previously solved model equivalence is obtained when there is no rewiring ($r=0$), the kernel is strictly the Albert-Barab\'asi preferential attachment ($\lambda=0$), both the node introduction rate and minimum degree are unit ($\Lambda=m=1$) and when the initial condition is described by two nodes connected with a single edge ($N_0=D_0=2$ and $N_\ell(0)=2\delta_{\ell,m}$).  From this description one can see that the model considered here is dynamically less-rich (the process of rewiring and of more general attachment is excluded) but considers a more general ``introduction regime" (the newly introduced nodes are a bit more general).\\

A common problem with the application of scale-free networks is that while it does match the form of many established structures, i.e. the World Wide Web (\cite{barabasi2000scale}), it does not properly match the evolution of such structures \cite{adamic2000power}.  The goal here is to produce explicit, time $t$, solutions which may be used in addressing questions which take place during the evolution of such network, e.g. synchronicity or ``worst-case" spread of disease on networks as considered in \cite{PhysRevLett.91.014101} and \cite{chowell28worst} respectively.  The ability to compare these time $t$ solutions with specific application data may allow for researchers to more accurately propose better models for the phenomena in question.\\

In Section \ref{gensolsec} I will go over the general recursive form of the solution found via integrating factors.  A few base cases, for $k=m,\,m+1,\,$ and $m+2$ are worked out in Section \ref{basecase}, and a proof by induction of the recognized form is given in Section \ref{Inductive}.  The paper ends with a discussion of the decay of dependence on initial conditions and a few ``tweaks" to and critiques of the model are suggested that may lead to better approximations.
\section{General Solution Form}\label{gensolsec}
For the sake of space we define the ``double-count" of all edges in the network at time $t$ by $D(t):=2\Lambda mt+D_0$, the total number of nodes as $N(t):=\Lambda t+N_0$, and let $R_k(t):=\pi_k(t)\Lambda m$.  Further note that

$$e^{\int R_k(t)dt}=D(t)^{\frac{k}{2}}=:H_k(t).$$

The general solution to $N_k(t)$ can thus be given recursively as
{\small \begin{eqnarray}
\dot{N}_k(t)+R_k(t)&=&\Lambda\delta_{k,m}+R_{k-1}(t)N_{k-1}(t)(1-\delta_{k,m}),\\
\dot{\left[N_k(t)H_k(t)\right]}&=&\left(\Lambda\delta_{k,m}+R_{k-1}(t)N_{k-1}(t)(1-\delta_{k,m})\right)H_k(t),
\end{eqnarray}
and finally
\begin{eqnarray}
N_k(t)&=&\frac{\int\left[\Lambda H_k(t)\delta_{k,m}+R_{k-1}(t)N_{k-1}(t)H_k(t)(1-\delta_{k,m})\right]dt+C_k}{H_k(t)},\\
&=&\frac{H_{2}(t)}{m(k+2)}\delta_{k,m}+\frac{\int R_{k-1}(t)N_{k-1}(t)H_k(t)(1-\delta_{k,m})dt+C_k}{H_k(t)},\label{gensol}
\end{eqnarray}
where the final step follows from the two observations (when $k\neq-2$)
\begin{eqnarray}
\nonumber\int H_k(t)dt&=&\int(2\Lambda mt+D_0)^{\frac{k}{2}}dt,\\
&=&\frac{(2\Lambda mt+D_0)^{\frac{k+2}{2}}}{\Lambda m(k+2)}=\frac{H_{k+2}(t)}{\Lambda m(k+2)},\label{Hprop1}
\end{eqnarray}
and
\begin{eqnarray}
\frac{H_{k+n}(t)}{H_{k}(t)}&=&\frac{(2\Lambda mt+D_0)^{\frac{k+n}{2}}}{(2\Lambda mt+D_0)^\frac{k}{2}}=H_n(t).\label{Hprop2}
\end{eqnarray}
The constants of integration, $C_k$, are defined by
\begin{eqnarray}
\nonumber C_k&=&N_k(0)-\frac{H_2(0)}{m(k+2)}\delta_{k,m}-\left.\frac{\int R_{k-1}(t)N_{k-1}(t)H_k(t)(1-\delta_{k,m})dt+C_k}{H_k(t)}\right|_{t=0}.
\end{eqnarray}}

These expressions have the general forms (proven later by induction)
{\begin{eqnarray}
N_k(t)=\frac{(m+1)\Gamma(k)H_2(t)}{\Gamma(k+3)}+\sum_{i=m}^{k}{k-1\choose i-1}\frac{C_{i}}{H_{i}(t)},\label{Nksol}
\end{eqnarray}}
with the constants of integration
{\small\begin{eqnarray}
C_{i}=H_i(0)\left(\frac{(-1)^{i-m+1}\Gamma(i)H_2(0)}{\Gamma(i-m+1)\Gamma(m+1)(i+2)}+\sum_{\ell=m}^{i}{i-1\choose \ell-1}(-1)^{i-\ell}N_\ell(0)\right).\label{Cisol}
\end{eqnarray}}
The typically assumed initial conditions are $N(0)=m+1$ and $D(0)=2m$ (i.e. $N_k(0)=\delta_{k,m}$). These expressions for $N_k(t)$ and $C_i$ condense to a closed form solution in terms of a hypergeometric function, ${}_2F_1$, (presented without derivation). Defining $G(t):=\frac{H_1(t)}{H_1(0)}=\sqrt{\frac{2\Lambda mt}{D_0}+1}$ the solution to $N_k(t)$ is 
\begin{eqnarray}
N_k(t)&=&\frac{(m+1)\Gamma(k)H_2(t)}{\Gamma(k+3)}-\frac{A_k(t)}{G^{m}(t)}+\frac{(G(t)-1)^{k-m}\Gamma(k)}{G^{k}(t)\Gamma(k-m+1)},
\end{eqnarray}
where
\begin{eqnarray}
A_k(t)&=&\frac{\Gamma(k){}_2F_1\left(m+2,m-k;m+3;\frac{1}{G(t)}\right)}{(m+2)\Gamma(m+1)\Gamma(k-m+1)}.
\end{eqnarray}
It is relatively straightforward to verify this result against that of Krapivsky and Redner (\cite{krapivsky2002finiteness}) using Equations \ref{Nksol} and \ref{Cisol} with $m=1$, $H_k(t)=(2t+2)^{\frac{k}{2}}$, and $N_\ell(0)=2\delta_{\ell,m}$ to arrive at
$$N_k(t)=\frac{4(t+1)}{k(k+1)(k+2)}+\frac{1}{(t+1)^{\frac{1}{2}}}\sum_{j=0}^{k-1}\frac{\Gamma(k)}{\Gamma(k-j)}\frac{(-1)^j(2j+4)}{j!(j+3)(t+1)^{\frac{j}{2}}}.$$\\

\section{Base Cases}\label{basecase}
In order to prove the previous solution expressions I employ a recursive/ inductive method.  The solution for the smallest degree, $m$, is calculated rather easily on its own with an integrating factor.  The degrees $k>m$ require the solution to the previous degree (recursion).  A pattern is found through inspection and then showed to hold in general in the inductive step in Section \ref{Inductive}.  A ``trick" is highlighted in the case of $N_{m+2}(t)$ which is employed to preserve the pattern.
\subsection{$N_{m+j}(t),\,\,j=0,1,2$}
In this Subsection I consider only the form of the state variable solution.  The form constants of integration are treated in the following Subsection.\\

\subsubsection{$j=0$}
The expression for $\dot{N}_m(t)$ is given by
$$\dot{N}_m(t)=\Lambda-R_m(t)N_m(t)=\Lambda-\Lambda m\pi_m(t)N_m(t).$$
The only additive term is the constant node introduction rate $\Lambda$ because each new node enters with degree $m$ and thus there may not exist a node of degree less than $m$ to ``feed into" $N_m(t)$.  The solution follows straightforwardly from Equation \ref{gensol}:
\begin{eqnarray}
N_m(t)&=&\frac{\Lambda\int H_m(t)dt+C_m}{H_m(t)},\\
&=&\frac{\Lambda \frac{H_{m+2}(t)}{\Lambda m(m+2)}+C_m}{H_m(t)},\\
&=&\frac{1}{m(m+2)}H_2(t)+\frac{C_m}{H_m(t)}.\label{j0}
\end{eqnarray}

\subsubsection{$j=1$}
The solution to $N_{m+1}(t)$ depends on $N_m(t)$, since a new node connecting to a node of degree $m$ creates a node of degree $m+1$.  Thus, again from Equation \ref{gensol}, we have a solution in the form:
\begin{eqnarray}
N_{m+1}(t)&=&\frac{\Lambda m\int\pi_m(t)N_m(t)H_{m+1}(t)dt+C_{m+1}}{H_{m+1}(t)},\\
&=&\frac{\Lambda m\int\frac{m}{H_2(t)}\left(\frac{H_2(t)}{m(m+2)}+\frac{C_m}{H_m(t)}\right)H_{m+1}(t)dt+C_{m+1}}{H_{m+1}(t)}.\label{j1gensol}
\end{eqnarray}
Considering just the integral we may employ the properties Equation \ref{Hprop1} and \ref{Hprop2} to write it as
\begin{eqnarray}
\int\left(\frac{H_{m+1}(t)}{m+2}+\frac{mC_m}{H_1(t)}\right)dt&=&\frac{H_{m+3}(t)}{\Lambda m(m+2)(m+3)}+\frac{C_m}{\Lambda}H_1(t).\label{j1int}
\end{eqnarray}
Plugging Equation \ref{j1int} into \ref{j1gensol} results in
\begin{eqnarray}
N_{m+1}(t)&=&\frac{1}{(m+2)(m+3)}H_2(t)+\frac{mC_m}{H_m(t)}+\frac{C_{m+1}}{H_{m+1}(t)}.\label{j1sol}
\end{eqnarray}

\subsubsection{$j=2$}
The solution to $N_{m+2}(t)$ follows in essentially identical fashion to that of $N_{m+1}(t)$ save for one ``tricky" step.  Within the integral term, as found in Equation \ref{j2gensol}, the term $m(m+1)C_m$ is written as $m(m+1)C_mH_{1-1}(t)$.  This will not alter the resultant value, but it does keep the form of $C_{m+i}$ consistent.  This way of {\em multiplying by 1} has to be employed for each consecutive ODE.  The solution via Equation \ref{gensol} is 
{\begin{eqnarray}
N_{m+2}(t)&=&\frac{\Lambda m\int\pi_{m+1}(t)N_{m+1}(t)H_{m+2}(t)dt+C_{m+2}}{H_{m+2}(t)},\\
&=&\frac{\Lambda m\int\left(\frac{(m+1)H_{m+2}(t)}{(m+2)(m+3)}+m(m+1)C_mH_{1-1}(t)+\frac{(m+1)C_{m+1}}{H_{1}(t)}\right)dt+C_{m+2}}{H_{m+2}(t)}.\label{j2gensol}
\end{eqnarray}}
This produces the solution
\begin{eqnarray}
N_{m+2}(t)=\frac{(m+1)}{(m+4)(m+3)(m+2)}H_2(t)+\frac{(m+1)mC_m}{2H_m(t)}+\frac{(m+1)C_{m+1}}{H_{m+1}(t)}+\frac{C_{m+2}}{H_{m+2}(t)}.\label{j2sol}
\end{eqnarray}

\subsubsection{The Pattern}
Inferring the pattern at this point is perhaps a touch premature.  Finding the solution to $N_{m+3}(t)$ makes the formulae evident, but for the sake of space I leave this to the reader and outline the pattern here using results from the $N_{m+3}(t)$ solution.  Each solution is composed of essentially two parts: a part that involves the term $H_2(t)$ and another that involves the constants of integration.\\

\paragraph{$H_2(t)$:}  The coefficients for the $H_2(t)$-term follow for $j=0,\,1,\,2,\,$ and $3$: $\frac{1}{m(m+2)},\,\frac{1}{(m+2)(m+3)},\,\frac{(m+1)}{(m+4)(m+3)(m+2)},\,$ and $\frac{(m+1)}{(m+5)(m+4)(m+3)}$.  The progression is clear when the first two terms are multiplied and divided by $m+1$ to give the general coefficient of the $H_2(t)$ term for $N_k(t)$ of 
\begin{equation}
\frac{m+1}{(k+2)(k+1)(k)}.
\end{equation}

\paragraph{$C_{m+i}$:} The coefficients for the $C_{m+i}$-term are each divided by the appropriate $H_{m+i}(t)$ term.  Furthermore, as is clearer from the solution of $N_{m+3}(t)$ there is also a term in the form of ${k-1\choose m+i-1}$ giving a final form for the coefficient of
\begin{equation}
{k-1\choose m+i-1}\frac{1}{H_{m+i}(t)}.
\end{equation}
Putting the two patterns together I infer a solution identical to Equation \ref{Nksol}.\\

\subsection{$C_{m+i},\,\,i=0,1,2$}
The constants of integration are found rather straightforwardly as
\begin{eqnarray*}
C_m&=&\left(N_m(0)-\frac{H_2(0)}{m(m+2)}\right)H_m(0),\\
C_{m+1}&=&\left(N_{m+1}(0)-mN_m(0)+\frac{H_2(0)}{m+3}\right)H_{m+1}(0),\\
C_{m+2}&=&\left(N_{m+2}(0)-(m+1)N_{m+1}(0)+\frac{m(m+1)}{2}N_m(0)-\frac{(m+1)H_2(0)}{2(m+4)}\right)H_{m+2}(0).
\end{eqnarray*}
The sign before the $H_2(0)$ terms is alternating and with a coefficient of ${m+i-1\choose i}\frac{1}{m(m+i+2)}$ for $C_{m+i}$.  The coefficient for the $N_{m+\ell}(0)$ terms is $(-1)^{i-\ell}{m+i-1\choose m+\ell-1}$ for $\ell=0,\, 1,\, ...,\, i$.  This gives an expression identical to Equation \ref{Cisol}.

\section{Inductive Step}\label{Inductive}
To prove that these are indeed the solutions for all $m+j,$ $j=0,\,1,\,2,\,...$ I will invoke induction.  Suppose that for some $j>2$ it is true that
$$N_{m+j}(t)=\frac{(m+1)\Gamma(m+j)H_2(t)}{\Gamma(m+j+3)}+\sum_{i=m}^{m+j}{m+j-1\choose i-1}\frac{C_{i}}{H_{i}(t)}.$$
By Equation \ref{gensol} we have that
$$N_{m+j+1}(t)=\frac{\Lambda m\int\pi_{m+j}(t)N_{m+j}(t)H_{m+j+1}(t)dt+C_{m+j+1}}{H_{m+j+1}(t)}.$$
Considering just the integral, a few cancellations gives the form
$$\frac{(m+1)\Gamma(m+j+1)}{\Gamma(m+j+3)}\int H_{m+j+1}(t)dt+(m+j)\sum_{i=m}^{m+j}{m+j-1\choose i-1}C_i\int H_{m+j-i-1}(t)dt.$$
By using the properties of $H$ in Equations \ref{Hprop1} and \ref{Hprop2} this resolves to
\begin{equation}
\frac{(m+1)\Gamma(m+j+1)}{\Lambda m\Gamma(m+j+4)}H_{m+j+3}(t)+\sum_{i=m}^{m+j}{m+j\choose i-1}C_i\frac{H_{m+j-i+1}(t)}{\Lambda m}.
\end{equation}
Placing this back into the general solution gives
\begin{equation}
N_{m+j+1}(t)=\frac{(m+1)\Gamma(m+j+1)}{\Gamma(m+j+4)}H_{2}(t)+\sum_{i=m}^{m+j}{m+j\choose i-1}\frac{C_i}{H_i(t)}+\frac{C_{m+j+1}}{H_{m+j+1}(t)},
\end{equation}
which is identical in form to Equation \ref{gensol}.  Thus, the general formula for $N_{k}(t)$ is correct.  Verifying $C_i$ follows straightforwardly from this proof.  The derivation of the single formula in terms of the hypergeometric function as in Equation \ref{gensol} is left to the reader and easily verified with a computer algebra system.

\section{Conclusion/ Discussion}
One may view the solution given in Equation \ref{Nksol} as containing                                                                                                                                                                                                                                                                                                                                                                                                                                                                                                                                                                                                                                                                                                                                                                                                                     two parts: the {\em core dynamics} and the {\em initial conditions}.  The initial conditions are contained in the series 
$$\sum_{i=m}^k{k-1\choose i-1}\frac{C_i}{H_i(t)},$$
or perhaps more illustratively in
\begin{equation}
\sum_{i=m}^k\frac{\Gamma(k)}{\Gamma(i)\Gamma(k-i+1)}C_i(2\Lambda mt+D_0)^{\frac{-i}{2}}.
\end{equation}
It is clear here that for the quantity $N_k(t)$ the dependence on initial conditions generated by all $N_j(t)$ with $j$ less than some $i$ decays on the order of $O(t^{-\frac{i}{2}})$.\\  

We may also consider $\frac{N_k(t)}{N(t)}$, the proportion of nodes that have degree $k$, and find that this both asymptotically agrees with previous findings of the degree distribution which do not consider transitory behavior and with those found by Krapivsky and Redner under appropriate restrictions.  Also, of mathematical interest, I've inadvertently deduced a seemingly novel identity for hypergeometric functions.  Given that $N(t)=\Lambda t+N_0=\sum_{k=m}^\infty N_k(t)$ one may conclude for $N_0=m+1$, $D_0=2m$, and $N_k(0)=\delta_{k,m}$ that
\begin{eqnarray}
\nonumber(1+\Lambda t)^\frac{m}{2}\frac{\Gamma(m)-m}{(m+2)\Gamma(m+1)}&=&\sum_{j=0}^\infty\frac{\Gamma(j+m)}{\Gamma(j+1)}{}_2F_1\left(m+2,-j;m+3;\frac{1}{\sqrt{\Lambda t+1}}\right).
\end{eqnarray}\\

Direct application of this time $t$ solution should be useful for two reasons: formation of and dynamics on complex networks. While the exact rules that particular real-world networks observe in their formation is an intractable key to their structure, one may compare such structures (data) with the solution given here to justify if this mechanism is applicable.  For dynamics on non-limiting case networks (e.g. diseases through a smaller population, spread of computer viruses on local networks, ideas spreading through {\em forming} social cliques, etc...) this solution may also be employed to describe the topology of such structures.  Results such as those found by Zhao {\em et. al}, on the fragility or scale-free networks to attacks on hubs, should be extended to include networks at are evolving over time \cite{PhysRevE.70.035101}.  It is possible, that since at time $t$ there is a smaller probability of hubs than in the limiting network (i.e. the tail is not fat enough yet) the network is in fact more resilient to cascading failures that the seemingly more robust limiting network.\\

There are a few flaws with considering these rate equations as a model for the A-B network growth.  In an arbitrarily small time interval from $t=0$ there has been mass shifted into compartments, $N_k(t)$, with arbitrarily large degree.  Indeed, the process only allows for $\lfloor2\Lambda mt\rfloor+D_0$ edges at any time $t$, and thus there should be a maximum degree available at any time.  Pastor-Satorras and Vespignani do this to some extent for general scale-free networks with the introduction of exponentially decaying tails induced on the degree distribution \cite{pastor2002epidemic}.  This was done in the interest of investigating the epidemic threshold for finite sized networks in order to find structures with noncritical spread despite the scale-free property.  This method however would be inadequate for an investigation into the structure of the networks as they evolve since the networks were taken to be static in their work.  Instead the ``cut-off" could be modeled implicitly with some sort of modified system that permits the correct number of ODEs to be acting at any time $t$.  A delay type differential equation system containing delays similar to $t-\lfloor\frac{k-N_0}{\Lambda}\rfloor$ and initial conditions like $N_k(t)=0$ for $t\in\left[-\lfloor\frac{k-N_0}{\Lambda}\rfloor,0\right)$ would perhaps do a better job at modeling the transitory dynamics of this system as they posses this sort of activation switch, i.e. they may only start to accept ``mass" once the delay exits the initial data.   This would force a finiteness that is otherwise lost in the traditional rate equation techniques.\\

Further work is also open in the gross generalization of this process.  One may consider the degree of a new node to be a {\em distributed parameter}.  In other words the degree a node is introduced with is some random variable $M$ with probability mass function $p_M$.  The concept of rewiring may be reintroduced to the model in addition to a suite of more general attachment kernels.  Under these generalizations it is very likely that the network may not retain an asymptotic distribution of $P(k)\propto k^{-\gamma}$, especially if the attachment kernel differs from proportional.  However, the true power of the A-B model should be in its flexibility to allow for generalizations that produces richer behavior and thus, farther reaching applications.  




\bibliographystyle{model1-num-names}
\bibliography{MyCollection}

\begin{thebibliography}{30}
\expandafter\ifx\csname natexlab\endcsname\relax\def\natexlab#1{#1}\fi
\providecommand{\bibinfo}[2]{#2}
\ifx\xfnm\relax \def\xfnm[#1]{\unskip,\space#1}\fi
\bibitem[{Barab{\'a}si and Albert(1999)}]{barabasi1999emergence}
\bibinfo{author}{A.~Barab{\'a}si}, \bibinfo{author}{R.~Albert},
\newblock \bibinfo{title}{{Emergence of scaling in random networks}},
\newblock \bibinfo{journal}{Science} \bibinfo{volume}{286}
  (\bibinfo{year}{1999}) \bibinfo{pages}{509}.
\bibitem[{Bianconi and Barab{\'a}si(2001)}]{bianconi2001bose}
\bibinfo{author}{G.~Bianconi}, \bibinfo{author}{A.~Barab{\'a}si},
\newblock \bibinfo{title}{{Bose-Einstein condensation in complex networks}},
\newblock \bibinfo{journal}{Physical Review Letters} \bibinfo{volume}{86}
  (\bibinfo{year}{2001}) \bibinfo{pages}{5632--5635}.
\bibitem[{Caldarelli(2007)}]{caldarelli2007scale}
\bibinfo{author}{G.~Caldarelli}, \bibinfo{title}{{Scale-Free Networks: Complex
  webs in nature and technology}}, \bibinfo{publisher}{Oxford University Press,
  USA}, \bibinfo{year}{2007}.
\bibitem[{Chowell and Castillo-Chavez(????)}]{chowell28worst}
\bibinfo{author}{G.~Chowell}, \bibinfo{author}{C.~Castillo-Chavez},
\newblock \bibinfo{title}{{Worst Case Scenarios and Epidemics}},
\newblock \bibinfo{journal}{Bioterrorism: mathematical modeling applications in
  homeland security} \bibinfo{volume}{28} (????).
\bibitem[{Christakis and Fowler(2007)}]{christakis2007spread}
\bibinfo{author}{N.~Christakis}, \bibinfo{author}{J.~Fowler},
\newblock \bibinfo{title}{{The spread of obesity in a large social network over
  32 years}},
\newblock \bibinfo{journal}{The New England Journal of Medicine}
  \bibinfo{volume}{357} (\bibinfo{year}{2007}) \bibinfo{pages}{370}.
\bibitem[{Lewis et~al.(2008)Lewis, Kaufman, Gonzalez, Wimmer, and
  Christakis}]{lewis2008tastes}
\bibinfo{author}{K.~Lewis}, \bibinfo{author}{J.~Kaufman},
  \bibinfo{author}{M.~Gonzalez}, \bibinfo{author}{A.~Wimmer},
  \bibinfo{author}{N.~Christakis},
\newblock \bibinfo{title}{{Tastes, ties, and time: A new social network dataset
  using Facebook. com}},
\newblock \bibinfo{journal}{Social Networks} \bibinfo{volume}{30}
  (\bibinfo{year}{2008}) \bibinfo{pages}{330--342}.
\bibitem[{Liu et~al.(2003)Liu, Lai, and Ye}]{PhysRevE.67.031911}
\bibinfo{author}{Z.~Liu}, \bibinfo{author}{Y.-C. Lai}, \bibinfo{author}{N.~Ye},
\newblock \bibinfo{title}{Propagation and immunization of infection on general
  networks with both homogeneous and heterogeneous components},
\newblock \bibinfo{journal}{Phys. Rev. E} \bibinfo{volume}{67}
  (\bibinfo{year}{2003}) \bibinfo{pages}{031911}.
\bibitem[{Motter et~al.(2002{\natexlab{a}})Motter, de~Moura, Lai, and
  Dasgupta}]{PhysRevE.65.065102}
\bibinfo{author}{A.~E. Motter}, \bibinfo{author}{A.~P.~S. de~Moura},
  \bibinfo{author}{Y.-C. Lai}, \bibinfo{author}{P.~Dasgupta},
\newblock \bibinfo{title}{Topology of the conceptual network of language},
\newblock \bibinfo{journal}{Phys. Rev. E} \bibinfo{volume}{65}
  (\bibinfo{year}{2002}{\natexlab{a}}) \bibinfo{pages}{065102}.
\bibitem[{Motter et~al.(2002{\natexlab{b}})Motter, Nishikawa, and
  Lai}]{PhysRevE.66.065103}
\bibinfo{author}{A.~E. Motter}, \bibinfo{author}{T.~Nishikawa},
  \bibinfo{author}{Y.-C. Lai},
\newblock \bibinfo{title}{Range-based attack on links in scale-free networks:
  Are long-range links responsible for the small-world phenomenon?},
\newblock \bibinfo{journal}{Phys. Rev. E} \bibinfo{volume}{66}
  (\bibinfo{year}{2002}{\natexlab{b}}) \bibinfo{pages}{065103}.
\bibitem[{Pastor-Satorras and Vespignani(2001)}]{pastor2001epidemic}
\bibinfo{author}{R.~Pastor-Satorras}, \bibinfo{author}{A.~Vespignani},
\newblock \bibinfo{title}{{Epidemic spreading in scale-free networks}},
\newblock \bibinfo{journal}{Physical review letters} \bibinfo{volume}{86}
  (\bibinfo{year}{2001}) \bibinfo{pages}{3200--3203}.
\bibitem[{Small et~al.(2007)Small, Walker, and Tse}]{small2007scale}
\bibinfo{author}{M.~Small}, \bibinfo{author}{D.~Walker},
  \bibinfo{author}{C.~Tse},
\newblock \bibinfo{title}{{Scale-free distribution of avian influenza
  outbreaks}},
\newblock \bibinfo{journal}{Physical review letters} \bibinfo{volume}{99}
  (\bibinfo{year}{2007}) \bibinfo{pages}{188702}.
\bibitem[{Price(1965)}]{price1965networks}
\bibinfo{author}{D.~Price},
\newblock \bibinfo{title}{{NETWORKS OF SCIENTIFIC PAPERS.}},
\newblock \bibinfo{journal}{Science (New York, NY)} \bibinfo{volume}{149}
  (\bibinfo{year}{1965}) \bibinfo{pages}{510}.
\bibitem[{Newman et~al.(2006)Newman, Barabasi, and Watts}]{newman2006structure}
\bibinfo{author}{M.~Newman}, \bibinfo{author}{A.~Barabasi},
  \bibinfo{author}{D.~Watts}, \bibinfo{title}{{The structure and dynamics of
  networks}}, \bibinfo{publisher}{Princeton Univ Pr}, \bibinfo{year}{2006}.
\bibitem[{Dorogovtsev et~al.(2000)Dorogovtsev, Mendes, and
  Samukhin}]{dorogovtsev2000structure}
\bibinfo{author}{S.~Dorogovtsev}, \bibinfo{author}{J.~Mendes},
  \bibinfo{author}{A.~Samukhin},
\newblock \bibinfo{title}{{Structure of growing networks with preferential
  linking}},
\newblock \bibinfo{journal}{Physical Review Letters} \bibinfo{volume}{85}
  (\bibinfo{year}{2000}) \bibinfo{pages}{4633--4636}.
\bibitem[{Krapivsky et~al.(2000)Krapivsky, Redner, and
  Leyvraz}]{krapivsky2000connectivity}
\bibinfo{author}{P.~Krapivsky}, \bibinfo{author}{S.~Redner},
  \bibinfo{author}{F.~Leyvraz},
\newblock \bibinfo{title}{{Connectivity of growing random networks}},
\newblock \bibinfo{journal}{Physical Review Letters} \bibinfo{volume}{85}
  (\bibinfo{year}{2000}) \bibinfo{pages}{4629--4632}.
\bibitem[{Krapivsky and Redner(2002)}]{krapivsky2002finiteness}
\bibinfo{author}{P.~Krapivsky}, \bibinfo{author}{S.~Redner},
\newblock \bibinfo{title}{{Finiteness and fluctuations in growing networks}},
\newblock \bibinfo{journal}{Journal of Physics A: Mathematical and General}
  \bibinfo{volume}{35} (\bibinfo{year}{2002}) \bibinfo{pages}{9517}.
\bibitem[{Krapivsky and Redner(2001)}]{krapivsky2001organization}
\bibinfo{author}{P.~Krapivsky}, \bibinfo{author}{S.~Redner},
\newblock \bibinfo{title}{{Organization of growing random networks}},
\newblock \bibinfo{journal}{Physical Review E} \bibinfo{volume}{63}
  (\bibinfo{year}{2001}) \bibinfo{pages}{066123}.
\bibitem[{Dorogovtsev and Mendes(2000)}]{dorogovtsev2000exactly}
\bibinfo{author}{S.~Dorogovtsev}, \bibinfo{author}{J.~Mendes},
\newblock \bibinfo{title}{{Exactly solvable small-world network}},
\newblock \bibinfo{journal}{EPL (Europhysics Letters)} \bibinfo{volume}{50}
  (\bibinfo{year}{2000}) \bibinfo{pages}{1}.
\bibitem[{Newman et~al.(2000)Newman, Moore, and Watts}]{newman2000mean}
\bibinfo{author}{M.~Newman}, \bibinfo{author}{C.~Moore},
  \bibinfo{author}{D.~Watts},
\newblock \bibinfo{title}{{Mean-field solution of the small-world network
  model}},
\newblock \bibinfo{journal}{Physical Review Letters} \bibinfo{volume}{84}
  (\bibinfo{year}{2000}) \bibinfo{pages}{3201--3204}.
\bibitem[{Wang et~al.(2003)Wang, Chakrabarti, Wang, and
  Faloutsos}]{wang2003epidemic}
\bibinfo{author}{Y.~Wang}, \bibinfo{author}{D.~Chakrabarti},
  \bibinfo{author}{C.~Wang}, \bibinfo{author}{C.~Faloutsos},
\newblock \bibinfo{title}{Epidemic spreading in real networks: An eigenvalue
  viewpoint},
\newblock \bibinfo{journal}{Reliable Distributed Systems, IEEE Symposium on}
  \bibinfo{volume}{0} (\bibinfo{year}{2003}) \bibinfo{pages}{25}.
\bibitem[{Valente(1996)}]{valente1996social}
\bibinfo{author}{T.~Valente},
\newblock \bibinfo{title}{{Social network thresholds in the diffusion of
  innovations}},
\newblock \bibinfo{journal}{Social Networks} \bibinfo{volume}{18}
  (\bibinfo{year}{1996}) \bibinfo{pages}{69--89}.
\bibitem[{Barab{\'a}si et~al.(1999)Barab{\'a}si, Albert, and
  Jeong}]{barabasi1999mean}
\bibinfo{author}{A.~Barab{\'a}si}, \bibinfo{author}{R.~Albert},
  \bibinfo{author}{H.~Jeong},
\newblock \bibinfo{title}{{Mean-field theory for scale-free random networks}},
\newblock \bibinfo{journal}{Physica A: Statistical Mechanics and its
  Applications} \bibinfo{volume}{272} (\bibinfo{year}{1999})
  \bibinfo{pages}{173--187}.
\bibitem[{Albert and Barab{\'a}si(2002)}]{albert2002statistical}
\bibinfo{author}{R.~Albert}, \bibinfo{author}{A.~Barab{\'a}si},
\newblock \bibinfo{title}{{Statistical mechanics of complex networks}},
\newblock \bibinfo{journal}{Reviews of modern physics} \bibinfo{volume}{74}
  (\bibinfo{year}{2002}) \bibinfo{pages}{47--97}.
\bibitem[{Boccaletti et~al.(2006)Boccaletti, Latora, Moreno, Chavez, and
  Hwang}]{boccaletti2006complex}
\bibinfo{author}{S.~Boccaletti}, \bibinfo{author}{V.~Latora},
  \bibinfo{author}{Y.~Moreno}, \bibinfo{author}{M.~Chavez},
  \bibinfo{author}{D.~Hwang},
\newblock \bibinfo{title}{{Complex networks: Structure and dynamics}},
\newblock \bibinfo{journal}{Physics Reports} \bibinfo{volume}{424}
  (\bibinfo{year}{2006}) \bibinfo{pages}{175--308}.
\bibitem[{Newman(2003)}]{newman2003structure}
\bibinfo{author}{M.~Newman},
\newblock \bibinfo{title}{{The structure and function of complex networks}},
\newblock \bibinfo{journal}{SIAM Review} \bibinfo{volume}{45}
  (\bibinfo{year}{2003}) \bibinfo{pages}{167--256}.
\bibitem[{Barab{\'a}si et~al.(2000)Barab{\'a}si, Albert, and
  Jeong}]{barabasi2000scale}
\bibinfo{author}{A.~Barab{\'a}si}, \bibinfo{author}{R.~Albert},
  \bibinfo{author}{H.~Jeong},
\newblock \bibinfo{title}{{Scale-free characteristics of random networks: the
  topology of the world-wide web}},
\newblock \bibinfo{journal}{Physica A: Statistical Mechanics and its
  Applications} \bibinfo{volume}{281} (\bibinfo{year}{2000})
  \bibinfo{pages}{69--77}.
\bibitem[{Adamic and Huberman(2000)}]{adamic2000power}
\bibinfo{author}{L.~Adamic}, \bibinfo{author}{B.~Huberman},
\newblock \bibinfo{title}{{Power-law distribution of the world wide web}},
\newblock \bibinfo{journal}{Science} \bibinfo{volume}{287}
  (\bibinfo{year}{2000}) \bibinfo{pages}{2115}.
\bibitem[{Nishikawa et~al.(2003)Nishikawa, Motter, Lai, and
  Hoppensteadt}]{PhysRevLett.91.014101}
\bibinfo{author}{T.~Nishikawa}, \bibinfo{author}{A.~E. Motter},
  \bibinfo{author}{Y.-C. Lai}, \bibinfo{author}{F.~C. Hoppensteadt},
\newblock \bibinfo{title}{Heterogeneity in oscillator networks: Are smaller
  worlds easier to synchronize?},
\newblock \bibinfo{journal}{Phys. Rev. Lett.} \bibinfo{volume}{91}
  (\bibinfo{year}{2003}) \bibinfo{pages}{014101}.
\bibitem[{Zhao et~al.(2004)Zhao, Park, and Lai}]{PhysRevE.70.035101}
\bibinfo{author}{L.~Zhao}, \bibinfo{author}{K.~Park}, \bibinfo{author}{Y.-C.
  Lai},
\newblock \bibinfo{title}{Attack vulnerability of scale-free networks due to
  cascading breakdown},
\newblock \bibinfo{journal}{Phys. Rev. E} \bibinfo{volume}{70}
  (\bibinfo{year}{2004}) \bibinfo{pages}{035101}.
\bibitem[{Pastor-Satorras and Vespignani(2002)}]{pastor2002epidemic}
\bibinfo{author}{R.~Pastor-Satorras}, \bibinfo{author}{A.~Vespignani},
\newblock \bibinfo{title}{{Epidemic dynamics in finite size scale-free
  networks}},
\newblock \bibinfo{journal}{Physical Review E} \bibinfo{volume}{65}
  (\bibinfo{year}{2002}) \bibinfo{pages}{035108}.

\end{thebibliography}







\end{document}